\documentclass[12pt]{article}
\newcommand{\be}{\begin{equation}}
\newcommand{\ee}{\end{equation}}
\newcommand{\bea}{\begin{eqnarray}}
\newcommand{\eea}{\end{eqnarray}}

\usepackage{dcolumn}% Align table columns on decimal point
\usepackage{bm}% bold math
\usepackage{amssymb,amsfonts,amsmath}
\usepackage{graphicx}% Include figure files
\usepackage[squaren]{SIunits}
\usepackage[super,numbers,sort&compress]{natbib}

\begin{document}
\baselineskip=30pt

\title{Griffiths singularity and magnetic phase diagram of La$_{1-x}$Ca$_x$CoO$_3$}
\author{Shiming Zhou$^\ast$, Yuqiao Guo, Laifa He, Jiyin Zhao, and Lei Shi\thanks{Authors to whom the correspondence should be
addressed. Electronic mail: zhousm@ustc.edu.cn; shil@ustc.edu.cn}\\
{\it Hefei National Laboratory for Physical Science at Microscale}\\
{\it University of Science and Technology of China}\\
{\it Hefei 230026, P. R. China}\\
}

\date{\today}
\maketitle

\newpage

\begin{abstract}
\baselineskip=24pt

Magnetic properties of La$_{1-x}$Ca$_{x}$CoO$_{3}$ (0.10 $\leq x
\leq$ 0.25) are systemically studied in this work. All the samples
exhibits the ferromagnetic state at low temperatures. However, their
inverse low-field magnetic susceptibilities shows a sharply downward
deviation from high-temperature Curie-Weiss paramagnetic behavior
well above the ferromagnetic transition temperature ($T_C$), which
indicates the presence of a ferromagnetic clustered state above
$T_C$. A detailed analysis on the susceptibilities reveals that the
short-range state in these Ca-doped samples can be well described as
the Griffiths phase. This characteristic is quite different from
those of the clustered states above $T_C$ recently reported in Sr-
and Ba-doped cobaltites, which are non-Griffith-like. It is proposed
that this difference possibly arises from the unique dependence of
magnetic interactions among Co$^{3+}$ ions on the size of the dopant
in the doped cobaltites. Based on these results, the magnetic
diagram of the Ca-doped cobaltites is established.

\end{abstract}

\section{Introduction}

Perovskite cobaltites La$_{1-x}$A$_{x}$CoO$_3$ ($A$ = Sr, Ba or Ca)
have recently attracted much attention since they exhibit various
intriguing physical properties such as magnetoresistance, large
thermoelectric effect, insulator-metal and spin-state
transitions.\cite{Briceno, Wang, Aarbogh, Tong, Radaelli,
Podlesnyak, Klie} The parent compound, LaCoO$_3$, has a nonmagnetic
insulating ground state with Co$^{3+}$ ions in a low spin
configuration. Upon warming, a paramagnetic (PM) insulating state
gradually develops above $\sim$ 90 K, where a spin-state transition
from low spin (LS) to higher spin state occurs. The partial
substitution of La$^{3+}$ ions by divalent earth-alkaline ions ($A$)
strongly affects the magnetic and transport properties of this
system due to the addition of holes into the lattice, creating
formally Co$^{4+}$ ions, and the structural changes because of the
different ionic radii of the substitutes.\cite{Podlesnyak2,
Podlesnyak3, Kuhns, Caciuffo, Wu, Giblin} A small doping can shift
the spin-state transition to low temperature and rapidly suppress
the nonmagnetic ground state. Consequently, ferromagnetic (FM)
correlations arising from the double-exchange (DE) coupling between
Co$^{3+}$-Co$^{4+}$ ions appear, which develop with doping and
result in a long-range FM ordered state above a critical doping
level ($x_C$).\cite{Podlesnyak2, Podlesnyak3} Moreover, various
experimental techniques such as nuclear magnetic resonance (NMR)
\cite{Kuhns}, small-angle neutron scattering \cite{Caciuffo, Wu},
and muon spin relaxation \cite{Giblin} have demonstrated that the
magnetic states of doped cobaltites at low temperatures are
inhomogeneous. Nanosized FM clusters were found by neutron
scattering \cite{Caciuffo, Wu} to form in the non-FM matrix, which
increases in density with doping and coalesce at $x_C$. NMR
measurements on the Sr-doped cobaltites revealed the coexistence of
FM regions, spin-glass regions, and hole-poor LS regions at low
temperatures.\cite{Kuhns}

Recently, magnetic phase inhomogeneity is also found at high
temperatures by the studies of small-angle neutron scattering and dc
susceptibilities on La$_{1-x}$Sr$_{x}$CoO$_3$, where a short-range
FM clustered state exists well above $T_C$.\cite{He} The existence
of FM clusters above $T_C$ is frequently reported in various FM
oxides such as manganites,\cite{Salamon, Deisenhofer, Jiang, Jiang2,
Pramanik, Zhou} layered cobaltites,\cite{Shimada} and spin-chain
compounds.\cite{Sampathkumaran} This often leads to Griffiths
singularity \cite{Griffiths}, which is originally proposed for
randomly diluted Ising ferromagnets. In the original model, the
nearest-neighbor exchange bonds with strength $J$ and 0 were argued
to be distributed randomly with probability $p$ and 1 - $p$,
respectively. For $p < p_{C}$ (percolation threshold), no long-range
FM order is established, while for $p \geq p_{C}$, the long-range FM
phase exists in a reduced $T_{C}(p)$ below the ordering temperature
of undiluted ferromagnet $T_{C}$($p$ = 1) known as Griffiths
temperature ($T_{G}$). The region $T_{C}(p) < T < T_{G}$, where the
system is characterized by the coexistence of FM clusters within the
globally PM phase, is refereed as the Griffiths phase. In doped
perovskite FM oxides, the quenched disorder induced by the
$A$-doping acts as random dilution. Thus, the Griffiths model can be
viewed as applicable to those oxides. Actually, for various doped
manganites including La$_{0.7}$Ca$_{0.3}$MnO$_{3}$,\cite{Salamon}
La$_{1-x}$Sr$_{x}$MnO$_{3}$ (0.07 $\leq x \leq$
0.16),\cite{Deisenhofer} and
La$_{0.73}$Ba$_{0.27}$MnO$_{3}$,\cite{Jiang} which exhibit many
similarities in the magnetic and transport properties to the doped
cobaltites such as hole-doping induced DE FM orderings and
percolation-type insulator-metal transitions,\cite{Aarbogh, Hoch}
the clustered phases above $T_{C}$ were reported to be well
described by the Griffiths phase. In those oxides, the Griffiths
phase is typically characterized by a sharply downward deviation
from the Curie-Weiss (CW) PM behavior in the low-field inverse
susceptibility ($\chi^{-1}(T)$) as the temperature approaches $T_C$
from above. However, for the Sr-doped cobaltites, it was found that
$\chi^{-1}(T)$ exhibits an upward deviation from the high-$T$ CW
behavior, which is in stark contrast to the predictions of the
Griffiths model and indicates the existence of non-Griffiths-like
clustered phase in those compounds \cite{He}. Similar
non-Griffiths-like clustered phase was also reported in
La$_{0.7}$Ba$_{0.3}$CoO$_3$.\cite{Huang} Those results seem to point
out that the FM clustered states above $T_{C}$ in the doped
cobaltites are quite different from those in the doped manganites.
However, the clear understanding on the formation of the unique
non-Griffiths-like phase is still lacking.

On the other hand, for doped cobaltites, many magnetic studies have
demonstrated that their magnetic properties strongly depend on the
ionic size of the dopant.\cite{Kriener, Kriener2, Phelan, Phelan2}
Specially, compared to the two cases doped by larger ions, i.e., $A$
= Sr or Ba, the Ca-doped crystals exhibit significant distinctions
in the magnetic behaviors. For example, for $A$ = Sr and Ba, $x_C$
is around 0.20,\cite{Kriener} while for $A$ = Ca, it is reported
that $x_C$ is much lower as about 0.05.\cite{Kriener2, Phelan}
Furthermore, at the same doping level, $T_C$ and the saturation
moment for Ca doped ones are somewhat lower and smaller than those
for $A$ = Sr and Ba, respectively.\cite{Kriener, Phelan} Recent
elastic neutron scattering revealed that in crystals with Sr and Ba,
Jahn-Teller (JT) spin polarons associated with intermediate-spin
(IS) state Co$^{3+}$ ions are present, whereas they are not detected
in those with Ca.\cite{Phelan2} Therefore, there are two interesting
issues to be addressed: (1) Whether is a short-range FM state
present above $T_C$ in the Ca-doped cobaltites, similar to the Sr-
and Ba-doped ones? (2) if present, whether is it a
non-Giriffiths-like or Griffiths-like phase? However, until now, to
the best of our knowledge, no study on the magnetic inhomogeneity
above $T_C$ for the Ca-doped cobaltites is carried out.

In this work, the magnetic properties of La$_{1-x}$Ca$_{x}$CoO$_{3}$
(LCCO) (0.10 $\leq x \leq$ 0.25) are systemically studied and the
above issues are focused on. We find that the magnetic
susceptibilities for all the Ca-doped samples deviate from the CW
law well above $T_C$, which indicates that a short-range FM state
also exists in this compound. However, quite different from the Sr-
and Ba-doped cobaltites, but similar to the doped manganites, the
deviation is sharply downward and the short-range FM state herein
can be well viewed as the Griffiths phase. The possible origin of
this important difference is discussed.

\section{Experimental Section}

Polycrystalline LCCO (0.10 $\leq x \leq$ 0.25) were prepared by
conventional solid-state reaction. The stoichiometric mixture of
La$_2$O$_3$, CaCO$_3$, and Co$_3$O$_4$ powders was well ground and
then calcined at 1000 and 1100 $\celsius$ for 24h with intermittent
grinding. The pellets pressed from the powders were sintered at 1200
$\celsius$ in the flowing oxygen for 48 h. The x-ray diffraction
(XRD) patterns were measured at room temperature on a Rigaku TTR-III
diffractometer using Cu K$\alpha$ radiation. The magnetic
measurements were carried out with a superconducting quantum
interference device magnetometer (Quantum Design MPMS XL-7).

\section{RESULTS AND DISCUSSION}

Room temperature XRD patterns of LCCO are shown in Fig. 1. All
diffraction peaks for each sample can be well indexed by the
perovskite structure without any impure phases. The samples with $x$
$\leq$ 0.18 has rhombohedral symmetry, while for $x$ = 0.25, it has
orthorhombic symmetry. Around $x$ = 0.20, there is a structural
change from rhombohedral to orthorhombic symmetry, as shown in the
inset of Figure 1. These results are in good agreement with the
previous structural studies on the Ca-doped
cobaltites.\cite{Kriener, Kriener2}

Figure 2 shows the temperature dependent field-cooling (FC)
magnetization under $H$ = 100 Oe for LCCO. Upon cooling, the
magnetization for all the samples shows a sharp rise, indicating the
FM ground state at low temperature. $T_C$, identified from the
minimum in $dM/dT$, increases from $\sim$ 52 to 151 K as $x$
increases from 0.10 to 0.25, as summarized in Figure 5.

The temperature dependence of $\chi^{-1}(T)$ under $H$ = 100 Oe for
all the samples are shown in Figure 3. The susceptibilities exhibit
a CW PM behavior at high temperatures, i.e. $\chi(T) =
C/(T-\Theta)$, where $C$ is Curie constant and $\theta$ is CW
temperature. The fitting $\theta$ and the effective magnetic moment
($\mu_{eff}$) obtained from the fitting $C$ are plotted in Figure 4.
Both the parameters show an increase with $x$. The increase in
$\theta$, together with the increase of $T_C$, indicates that the FM
interactions are enhanced with doping. For doped cobaltites, the
theoretical value of $\mu_{eff}$ is given by $\mu_{eff}$ =
$g$$\sqrt{(1-x)S^{3+}(S^{3+}+1) + xS^{4+}(S^{4+}+1)}$, where
$S^{3+}$ and $S^{4+}$ are the spin value of Co$^{3+}$ and Co$^{4+}$
ions, respectively, and $g$ is Land\'{e} g factor. Both Co ions have
three possible spin states, \emph{i.e.}, LS, IS, and high-spin (HS)
states. For Co$^{3+}$ ions, it is widely accepted to be in IS state
in doped cobaltites, while for Co$^{4+}$ ions the spin state is more
controversial.\cite{Wu2, Burley, Tsubouchi} Assuming the spin state
of Co$^{4+}$ ions are in LS, IS, and HS ones, respectively, we can
calculate the theoretical $\mu_{eff}$ as a function of $x$, which
are shown in Figure 4(a). It is clear that the IS case is most in
accord with the experimental result. Therefore, both Co ions are in
IS states for those Ca-doped samples in the high-$T$ PM state.

Upon cooling, one can see that $\chi^{-1}(T)$ for LCCO shows a
deviation from the CW law well above $T_C$, which strongly suggests
that a short-range FM state exists before the long-range FM
transition in those compounds. The $x$-dependence of $T_G$, i.e.,
the temperature below which $\chi^{-1}(T)$ starts to deviate, is
plotted in Figure 5, which shows a slight increase from 161 to 181 K
with doping. This result together with those recently reported in
the Sr- and Ba-doped cobaltites \cite{He, Huang} imply that the
preformation of the FM clustered state above $T_C$ should be a
common phenomenon in hole-doped LaCoO$_3$. However, in contrast with
the two cases doped by larger ions, where the deviation in
$\chi^{-1}(T)$ is upward, the Ca-doped samples exhibit the sharply
downward deviations in $\chi^{-1}(T)$. The downturn, similar to
those reported in the doped manganites,\cite{Salamon, Deisenhofer,
Jiang} is a typical characteristic of the Griffiths phase. In the
Griffiths phase, the downward deviation in low-field $\chi^{-1}(T)$
is proposed to originate from the enhanced $\chi(T)$ due to the
contribution from the FM clusters above $T_C$, and can be gradually
suppressed with increasing $H$ due to polarization of spins outside
the clusters.\cite{Jiang, Jiang2, Pramanik, He} To verify the latter
feature, we have further measured the magnetizations under $H$ = 500
and 5k Oe for those compounds and plotted the corresponding
$\chi^{-1}(T)$ in Figure 3, too. It is clearly seen that the
deviation is markedly suppressed indeed under the larger magnetic
fields for all the samples, which supports the presence of the
Griffiths singularity in LCCO.

According to the model of Griffiths phase, the system exhibits
neither a pure PM behavior nor a long-range FM order in the
Griffiths phase regime.\cite{Griffiths, Jiang, Pramanik, Jiang2}
Consequently, the system response is dominated by the largest
magnetic cluster/correlated volume, which will give rise to a
characteristic $T$-dependence for the low-field susceptibility by
the following power law: $\chi^{-1}(T) \propto (T - T^{R}_{C})^{1 -
\lambda}$, where $\lambda$ is the susceptibility exponent and 0 $<
\lambda <$ 1.\cite{Jiang, Pramanik, Jiang2} In order to further
confirm the Griffiths singularity in LCCO, we have fitted
$\chi^{-1}(T)$ under $H$ = 100 Oe by the above law for all the
samples. It is pertinent to note that an incorrect value of
$T^{R}_{C}$ in this formula can lead to unphysical fitting and
erroneous determination of $\lambda$. To estimate $\lambda$
accurately, we have followed the approach by Jiang et
al.\cite{Jiang2}, where $T^{R}_{C}$ is correctly given as fitting
the data in the pure PM region above $T_{G}$ to this law yields a
value of $\lambda$ close to zero. In that approach, $T^{R}_{C}$ is
essentially equivalent to the value of $\theta$. The
double-logarithmic plots of $\chi^{-1}$ against reduced temperature
$t_{m} = (T - T^{R}_{C})/T^{R}_{C}$ for all the samples are
displayed in the inset of Figure 3. From the slope of the fitted
straight line in the Griffiths phase regime, the exponent $\lambda$
for LCCO is obtained, which is plotted in Figure 4(c). For all the
samples, the values of $\lambda$ are less than unity, well
consistent with the expectation from the Griffiths phase model.
Moreover, it is found that the exponent $\lambda$ decreases with the
increase of $x$. Since $\lambda$ signifies the deviation from the CW
behavior and higher its value the stronger is the
deviation,\cite{Pramanik} this decrease implies that the Griffiths
phase is weakened as $x$ increases, which well agrees with the fact
that the temperature range of the Griffiths phase decreases upon
doping as shown in the magnetic phase diagram of LCCO (see Figure
5). This evolution of the Griffiths phase with the composition is
very similar to those reported in the doped manganite
La$_{1-x}$Sr$_{x}$MnO$_{3}$ \cite{Deisenhofer} and
La$_{1-x}$Ba$_{x}$MnO$_{3}$ \cite{Jiang} and can be comparable with
the $T-p$ phase diagram from the Griffiths model.\cite{Deisenhofer,
Jiang} These features strongly indicate that the clustered state in
LCCO can be well described by the Griffiths phase.

Our magnetic studies unequivocally reveal that the FM clustered
state above $T_C$ in the Ca-doped cobaltites is significantly
different from those reported in the Sr- and Ba-doped ones. Now, let
us to discuss the possible origin of this important difference.
Usually, the quenched disorder or the competition between magnetic
interactions are argued to be the fundamental ingredient in the
onset of the Griffiths phase.\cite{Pramanik, Jiang, Jiang2} For
example, in La$_{1-x}$Ba$_x$MnO$_3$, the quenched disorder arising
from the size variance of La/Ba atoms was reported to be responsible
for the development of the Griffiths phase,\cite{Jiang} while in
(La$_{1-y}$Pr$_y$)$_{0.7}$Ca$_{0.3}$Mn$^{16/18}$O$_3$, a close
relationship between the FM-antiferromagnetic (AFM) phase
competition and the nucleation of the Griffiths phase was observed,
where the Griffiths phase appears as the FM phase dominates and
disappears as the AFM phase dominates over the FM one.\cite{Jiang2}
Recently, we have found that a size-induced transition from
non-Griffiths to Griffiths phase exists in
Sm$_{0.5}$Sr$_{0.5}$MnO$_{3}$ nanoparticles, which is proposed due
to the strong suppression of the AFM interactions above $T_C$ by the
size reduction.\cite{Zhou} For the doped cobaltites, since the
samples with $A$ = Sr or Ba have a relatively larger quenched
disorder because of the larger difference in the radii between La
and $A$ ions, the quenched disorder seems not the origin of the
Griffiths phase in the Ca-doped samples. For the Sr-and Ba-doped
ones, the observed upward deviation in $\chi^{-1}(T)$ from the CW
law implies that $\chi(T)$ is reduced to be lower than the value
expected from the pure PM behavior. This reduction is most probably
due to the presence of AFM interactions. Alternatively, He et al.
speculated that the interactions between the antiparallel alignment
of neighboring FM clusters could be AFM.\cite{He} We note that in
the doped cobaltites the DE coupling between Co$^{3+}$-Co$^{4+}$
ions is FM, while the superexchange interactions among IS Co$^{3+}$
ions can be AFM or FM, depending on whether the splitting of $e_{g}$
orbitals, \emph{i.e.}, JT distortion, happen or not.\cite{Fuchs,
Zhou2, Zhou3} In other words, the $e_{g}$ orbital ordering of
Co$^{3+}$ ions with JT distortion yields AFM spin correlations,
whereas the suppression of the JT distortion will lead the orbitals
to favor FM superexchange. For La$_{1-x}$A$_{x}$CoO$_{3}$, the JT
distortion is strongly dependent on the ionic size of the A-site
dopant. Structural studies by the neutron diffractions have
disclosed that the JT distortion associated with IS Co$^{3+}$ ions
are present in crystals with Sr and Ba but suppressed in those with
Ca.\cite{Phelan, Phelan2} This gives a clue to argue that the
interactions between Co$^{3+}$ ions are AFM in the crystals doped by
Sr and Ba but are FM in the case by Ca. Practically, recent elastic
neutron scattering on the doped cobaltites indeed revealed that the
AFM ordered state is observed only for $A$ = Sr and Ba but not for A
= Ca.\cite{Yu} Therefore, we propose that for $A$ = Sr and Ba the
presence of the AFM correlations from the Co$^{3+}$ ions with the JT
distortion results in the upward deviation in $\chi^{-1}(T)$ and
hence the non-Griffiths-like phase. However, for $A$ = Ca, the AFM
interactions are suppressed due to the absence of the JT distortion,
which promotes the appearance of the Girffiths phase.

\section {Conclusions}

In conclusion, the magnetic studies on La$_{1-x}$Ca$_{x}$CoO$_{3}$
reveal that the FM clustered state exists above $T_C$ in this
compound. Moreover, it is found that this state has the basic
characteristics of the Girffiths phase, which is quite different
from those recently reported in the Sr- and Ba-doped cobaltites
where they are non-Girffiths-like. It is proposed that this
difference possibly arises from the quite distinct magnetic
interactions between Co$^{3+}$ ions in those doped cobaltites. On
the base of these results, we establish the magnetic diagram of the
Ca-doped cobaltites.

\section {Acknowledgement}

This project was financially supported by the National Science
Foundation of China (Grant No.10904135), the National Basic Research
Program of China (973 program, 2012CB927402 and 2009CB939901), and
the Foundation for the Excellent Youth Scholars of Anhui Province of
China (No.2010SQRL007ZD).

\newpage

\begin{figure}
\center
\includegraphics[width=16cm]{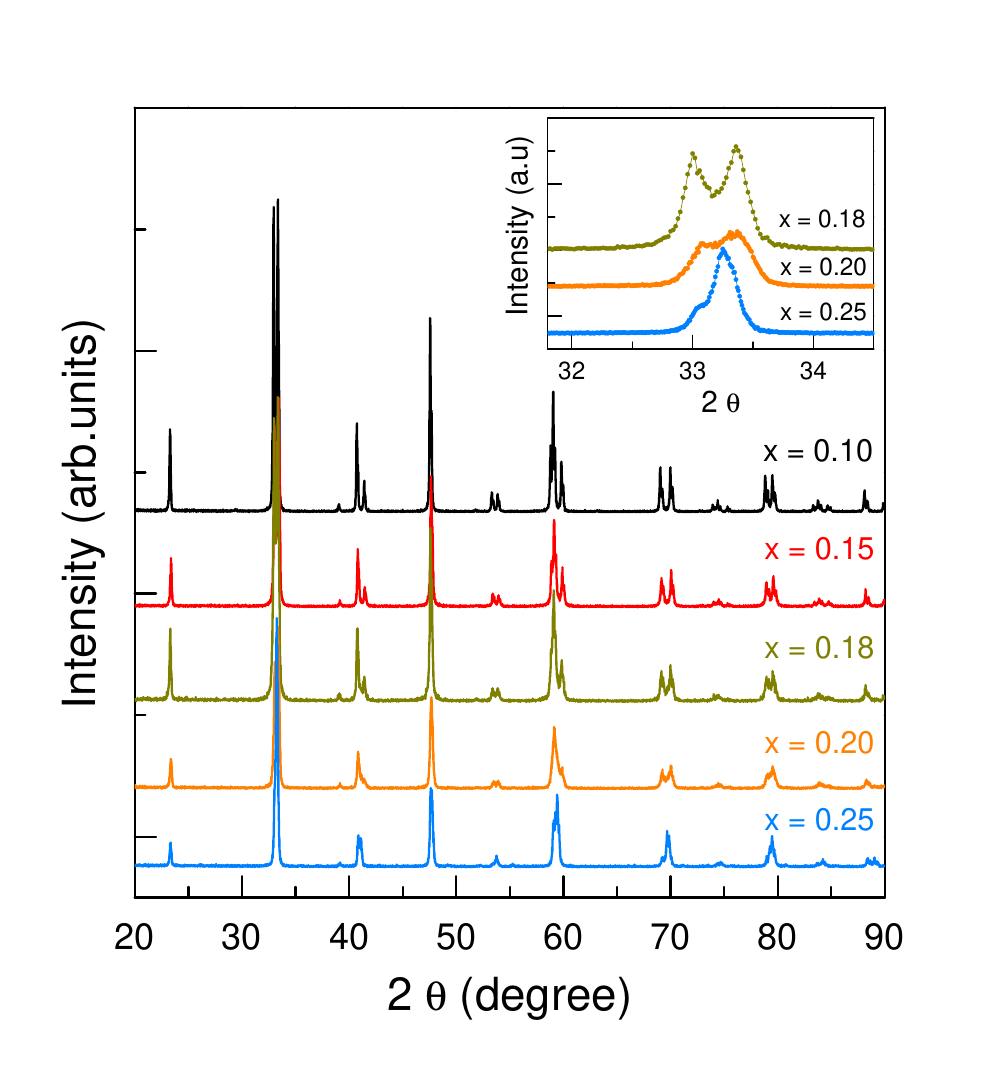}
\caption{(Color online) Room-temperature XRD patterns for
La$_{1-x}$Ca$_{x}$CoO$_{3}$ (0.10 $\leq x \leq$ 0.25). The insets
shows the structural change around $x$ = 0.20.} \label{Figure 1}
\end{figure}

\begin{figure}
\center
\includegraphics[width=16.8cm]{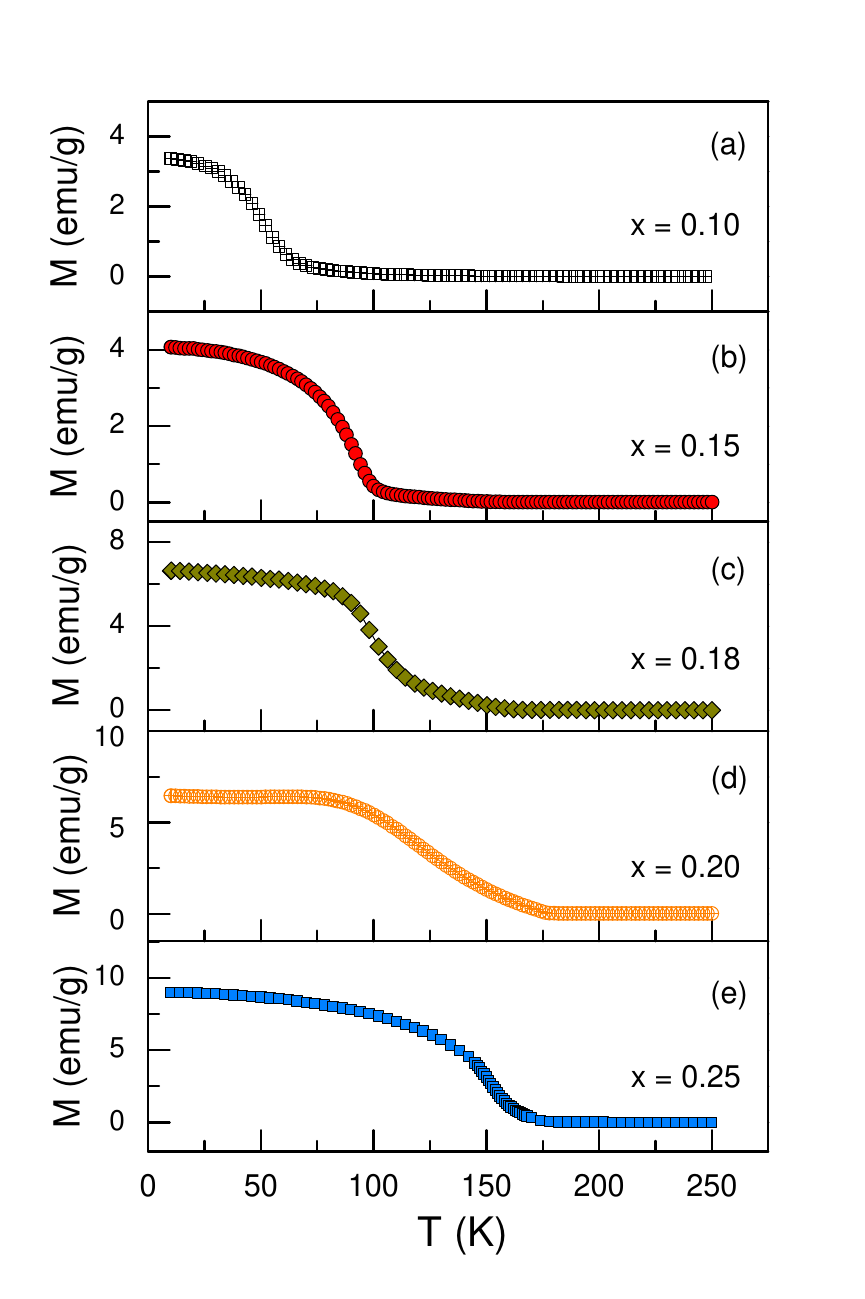}
\caption{(Color online) Temperature dependence of the FC
magnetization under $H$ = 100 Oe for La$_{1-x}$Ca$_{x}$CoO$_{3}$
(0.10 $\leq x \leq$ 0.25).} \label{Figure 2}
\end{figure}

\begin{figure}
\center
\includegraphics[width=8.4cm]{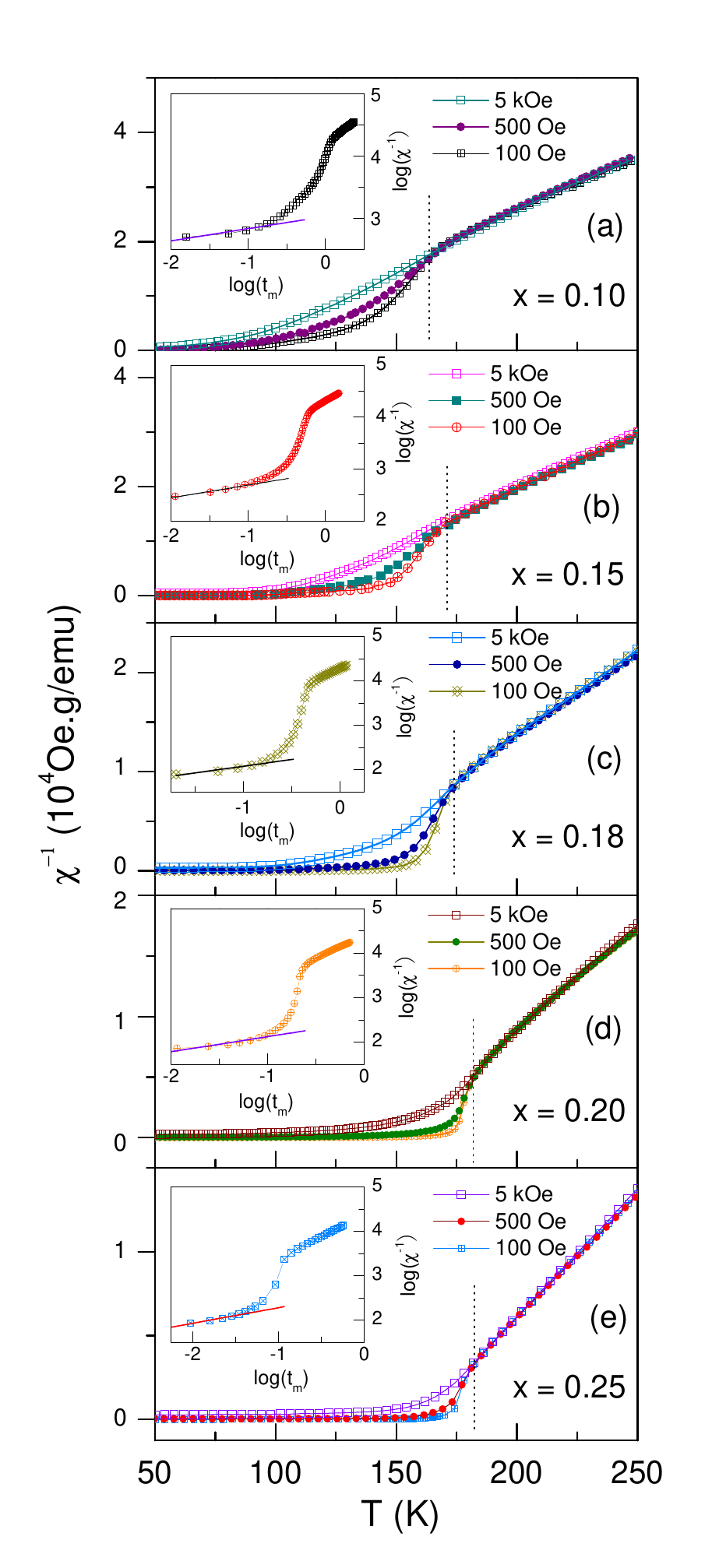}
\caption{(Color online) Temperature dependence of the inverse
susceptibilities under different magnetic fields for
La$_{1-x}$Ca$_{x}$CoO$_{3}$ (0.10 $\leq x \leq$ 0.25). The insets
show log($\chi^{-1}(T)$) vs log($t_m$) plots under $H$ = 100 Oe,
where the solid lines are the linear fittings.} \label{Figure 3}
\end{figure}

\begin{figure}
\center
\includegraphics[width=16.8cm]{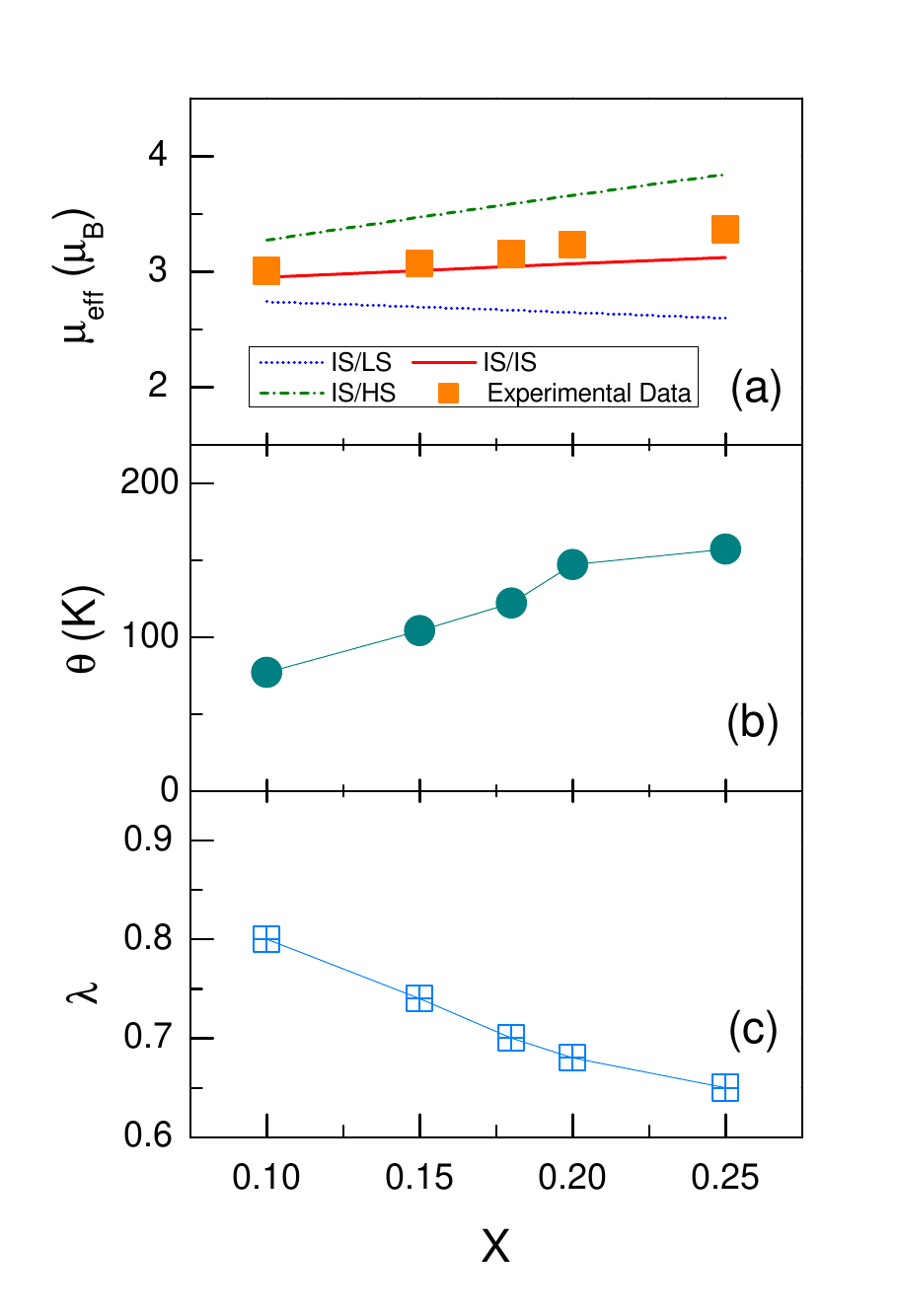}
\caption{(Color online) Ca-doping dependence of $\mu_{eff}$ (a),
$\theta$ (b), and $\lambda$ (c). The three lines in (a) are the
theoretically effective magnetic moment with the spin states of
Co$^{3+}$/Co$^{4+}$ ions in IS/LS, IS/IS, and IS/HS states,
respectively.} \label{Figure 4}
\end{figure}

\begin{figure}
\center
\includegraphics[width=16.8cm]{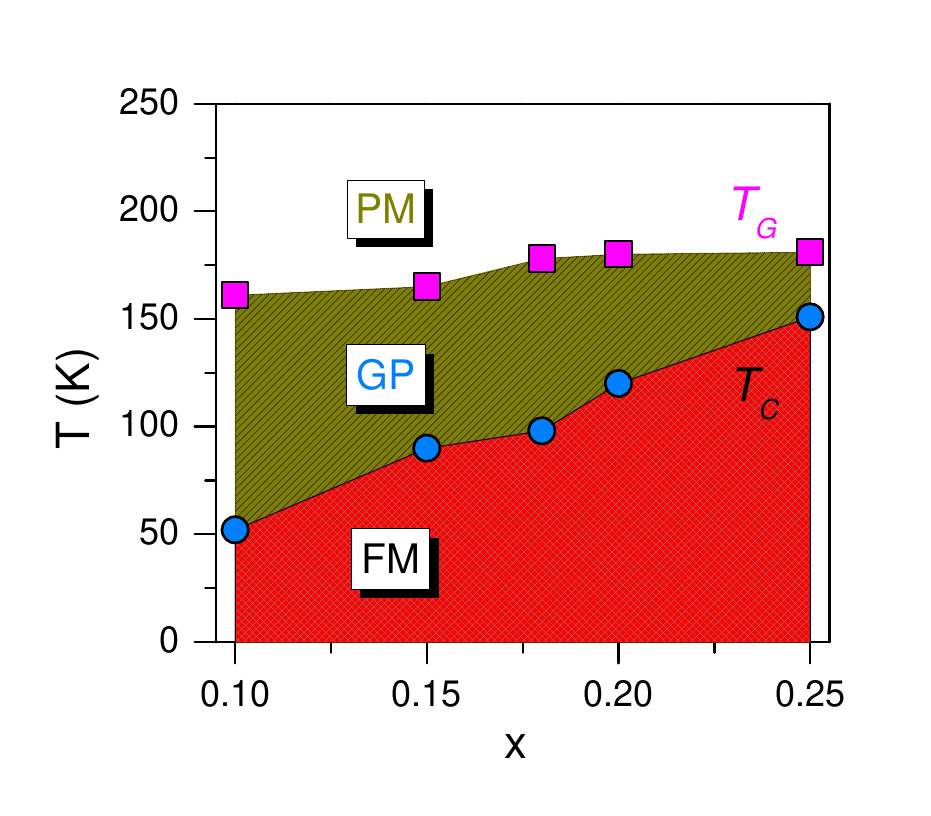}
\caption{(Color online) Magnetic phase diagram of
La$_{1-x}$Ca$_{x}$CoO$_{3}$ (0.10 $\leq x \leq$ 0.25). PM, GP, and
FM denote the paramagnetic, Griffiths, and ferromagnetic phase,
respectively.} \label{Figure 5}
\end{figure}

\end{document}